\documentclass[12pt]{article}
\usepackage{ulem}
\usepackage{authblk}
\usepackage{graphicx}
\usepackage{float}

\title{Terahertz multi-heterodyne spectroscopy using laser frequency combs}

\author[1,$\dagger$,*]{Yang Yang}
\author[1,$\dagger$,*]{David Burghoff}
\author[2]{Darren J. Hayton}
\author[2,3]{Jian-Rong Gao}
\author[4]{John L. Reno}
\author[1]{Qing Hu}

\affil[1]{Department of Electrical Engineering and Computer Science, Research Laboratory of Electronics, Massachusetts Institute of Technology, Cambridge, Massachusetts, 02139}
\affil[2]{SRON Netherlands Institute for Space Research, 9747 AD, Groningen, The Netherlands}
\affil[3]{Kavli Institute of NanoScience, Delft University of Technology, Lorentzweg 1, 2628 CJ Delft, The Netherlands}
\affil[4]{Center for Integrated Nanotechnology, Sandia National Laboratories, Albuquerque, New Mexico, 87123}

\affil[$\dagger$]{These authors contributed equally to this work.}
\affil[*]{Corresponding author: yang$\_$y@mit.edu, burghoff@mit.edu}

\date{}

\begin{document}

\maketitle

\begin{abstract}
	Frequency combs based on terahertz quantum cascade lasers feature broadband coverage and high output powers in a compact package, making them an attractive option for broadband spectroscopy. Here, we demonstrate the first multi-heterodyne spectroscopy using two terahertz quantum cascade laser combs. With just 100 $\mu$s of integration time, we achieve peak signal-to-noise ratios exceeding 60 dB and a spectral coverage greater than 250 GHz centered at 2.8 THz. Even with room-temperature detectors we are able to achieve peak signal-to-noise ratios of 50 dB, and as a proof-of-principle we use these combs to measure the broadband transmission spectrum of etalon samples. Finally, we show that with proper signal processing, it is possible to extend the multi-heterodyne spectroscopy to quantum cascade laser combs operating in pulsed mode, greatly expanding the range of quantum cascade lasers that could be suitable for these techniques.
\end{abstract}


Terahertz radiation, light whose frequency lies in the 1 to 10 THz range, is of great importance for spectroscopy since many molecules have strong rotational and vibrational resonances in this frequency range. As a result, much effort has been spent developing terahertz spectroscopic techniques to address the trade-off between detection bandwidth, frequency resolution, and acquisition time.  Fourier domain techniques such as Fourier Transform infrared spectroscopy (FTIR) and terahertz time-domain spectroscopy (THz-TDS) are intrinsically broadband \cite{Finneran2015}, but their average power is quite low; moreover, their operations typically require a mechanical moving stage to achieve maximum signal-to-noise ratios. Impressive results have also been achieved with electronic THz sources based on cascaded frequency multiplication \cite{Drouin2005,Pearson2011}, but these sources suffer parasitic roll-offs and lower power levels especially at higher frequencies. Tunable laser sources based on high-power terahertz quantum cascade lasers (QCLs) \cite{qin_tuning_2009} are another alternative: by tuning the source, one can reconstruct sample's absorption feature over the tuning span \cite{bartalini_frequency-comb-assisted_2014} or conduct heterodyne detection \cite{ren_high-resolution_2011}. These approaches can achieve high frequency resolution, but are usually narrowband, owing to the limited tunability of the source. THz sources based on intra-cavity difference-frequency generation \cite{lu_widely_2014,jiang_spectroscopic_2016} can be broadly tuned, but only offer microwatt power levels when operated in continuous wave mode.

\begin{figure}[H]
	\centering
	\includegraphics[width=\linewidth]{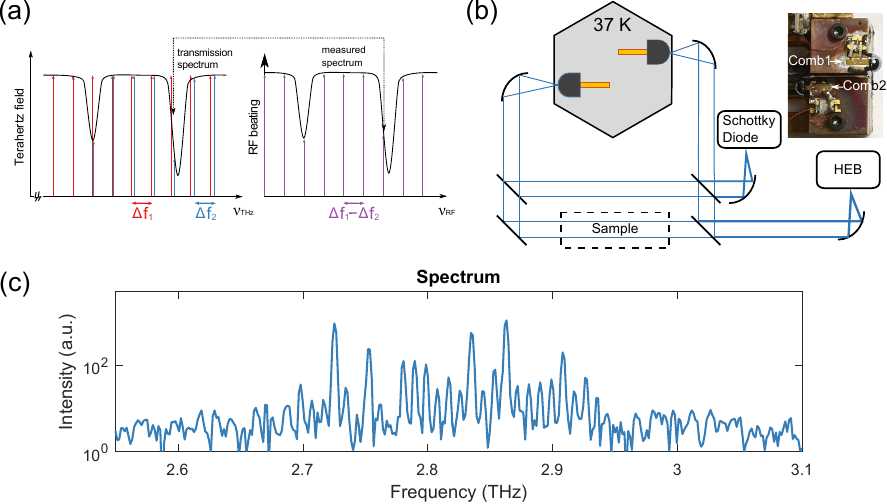}
	\caption{(a) Multi-heterodyne spectroscopy using two frequency combs with slightly different repetition rates. (b) Experimental set-up. Inset shows real laser frequency combs on the copper mount, both of which are silicon lens-coupled. (c) Lasing spectrum of one device under comb operation.}
	\label{fig:1}
\end{figure}

In contrast, multi-heterodyne spectroscopy based on two frequency combs \cite{schiller_spectrometry_2002,keilmann_time-domain_2004}, also known as dual-comb spectroscopy, offers an elegant way of conducting broadband spectroscopy, featuring broad spectral coverage, high frequency resolution, and high signal-to-noise ratios obtained over short acquisition times, all without mechanical moving parts \cite{coddington_coherent_2008,newbury_sensitivity_2010}. The principle is illustrated in Figure \ref{fig:1}(a): by beating two combs with slightly different repetition rates onto a single fast detector, one can measure a multitude of down-converted radio frequency (RF) beatnotes, each of which corresponds uniquely to the optical frequency beatings between adjacent lines from different combs. If one comb is shined through a sample, the sample’s absorption information at optical frequencies is encoded in the RF spectrum.

Traditionally, frequency combs have been generated in the THz range by downconversion of ultrafast laser pulses, which are used to form a time-domain THz pulse with a well-defined phase relation \cite{Tonouchi2007, Yasui2015}.  More recently, there has been great interest in THz combs based on quantum cascade lasers, which are generated by nonlinearities in low-dispersion cavities \cite{Hugi2012,Khurgin2014}. Such combs can be generated by deliberate dispersion engineering of the laser cavity \cite{burghoff_terahertz_2014} or by utilizing the naturally-low dispersion of the gain medium at a particular bias \cite{Wienold2014,rosch_octave-spanning_2015,Li2015}. From a spectroscopy perspective, QCL-based frequency combs offer compactness, continuous-wave (CW) operation, and high output powers. Moreover, their non-pulsed nature making them less prone to the detector saturation and gives them a larger dynamic range \cite{wang_high-resolution_2014, villares_dual-comb_2014}. Here, we present the first demonstration of using THz QCL combs to perform multi-heterodyne spectroscopy.

Figure \ref{fig:1}(b) shows a simplified experimental set-up used for this demonstration. For simplicity, only the optical path is illustrated here. Using integrated dispersion compensation, two THz QCL combs were fabricated using a gain medium similar to the one in Ref. \cite{rosch_octave-spanning_2015}. Both lasers were lens-coupled and had sub-milliwatt output powers at 37 K; when biased into the comb regime, their lasing spectra covered approximately 250 GHz at 2.8 THz (shown in Fig. \ref{fig:1}(c)). To minimize their environmental differences, both devices were mounted inside the same pulsed-tube cryocooler. To account for amplitude fluctuations, a balanced detection scheme was employed, using one superconducting hot-electron bolometer (HEB) mixer \cite{Zhang2010} as the signal detector and one Schottky mixer (Virginia Diode's WR-0.34HM) as the reference detector. Naturally, the two detectors were quite different in terms of sensitivity and dynamic range—each detector must be both fast and sensitive, a challenging requirement in the terahertz range—and so some differences will be evident in the resulting measured spectra. In particular, the Schottky mixer is operated at room temperature and is less sensitive, whereas the HEB is helium-cooled and is more sensitive, but is also significantly less linear.

\begin{figure}[H]
	\centering
	\includegraphics[width=\linewidth]{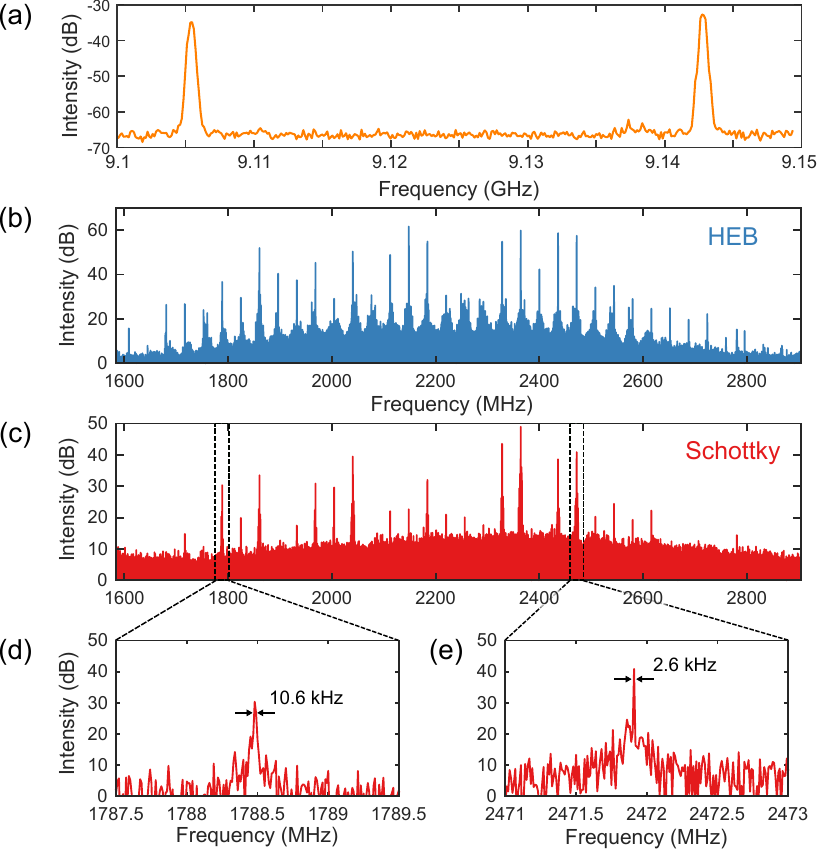}
	\caption{(a) Repetition rates of two combs detected from a bias tee. The two repetition rate beatnotes are located at around 9.1 GHz with a difference of about 36 MHz. (b) Multi-heterodyne signal on the HEB with 100 $\mu$s acquisition time. The detector noise floor is at 0 dB. (c) Multi-heterodyne signal on the Schottky mixer with 100 $\mu$s acquisition time. (d), (e) Zoomed-in view of multi-heterodyne teeth from Schottky mixer at 1788.5 MHz and 2472 MHz, respectively. The FWHM of the 1788.5 MHz tooth is 10.6 kHz and the FWHM of the 2472 MHz one is 2.6 kHz.}
	\label{fig:2}
\end{figure}

Both QCLs were biased into a comb regime, and the repetition rate beatnotes were detected from them using a bias tee and are shown in Fig. \ref{fig:2}(a). The free-running combs featured repetition rates around 9.1 GHz and were separated by a 36 MHz difference, i.e. $\Delta f_2- \Delta f_1$=36 MHz. At the same time, we also detected a multi-heterodyne RF signal centered at 2.2 GHz from both the HEB and the Schottky mixer, indicating that these two combs’ offset frequency differed by about 2.2 GHz, i.e. $f_{\textrm{ceo,2}}- f_{\textrm{ceo,1}}$=2.2 GHz. To avoid the need for a high-speed data recorder, the multi-heterodyne signal was downconverted into the oscilloscope’s bandwidth by IQ-demodulating with a synthesizer of frequency $f_{LO}$=2.364 GHz. Both the in-phase and in-quadrature signals were then recorded with a fast oscilloscope. 

The downconverted multi-heterodyne signals were recorded for a duration of 100 $\mu$s and are shown in Figs. \ref{fig:2}(b) and \ref{fig:2}(c). The signal from the HEB was used to generate a phase and timing correction signal \cite{itqw}, and this signal was used to correct both interferograms \cite{Roy2012,ideguchi_adaptive_2014,Yasui2015}. Note that while this procedure generates only two correction frequencies, because the radiation has the highly coherent structure of a comb, only two frequency parameters are needed to correct \textit{all} of the multi-heterodyne lines. For example, Figs. \ref{fig:2}(d) and \ref{fig:2}(e) show two multi-heterodyne teeth from the Schottky mixer, located at 1788.5 MHz and 2472 MHz, which have full-width half-maximum (FWHM) linewidths of 10.6 kHz and 2.6 kHz, respectively. Both linewidths are at the Fourier uncertainty limit, implying that our correction procedure has removed most of the phase and timing errors. (Note also that because phase correction deletes mutual phase fluctuations, limitations in the linewidth imposed by quantum fluctuations \cite{Vitiello2012,Bartalini2010} do not pertain here.) The leftover multiplicative noise after the phase and timing correction contributes to the noise floor of both multi-heterodyne signals, forming a pedestal as shown in Fig. \ref{fig:2}(b) and Fig. \ref{fig:2}(c). 

With an acquisition time of 100 $\mu$s, the peak signal-to-noise ratio (SNR) from the HEB is higher than 60 dB and the multi-heterodyne signal spans 1.08 GHz with 30 distinguishable teeth, corresponding to an optical spectrum coverage greater than 250 GHz at 2.8 THz. The signal from the Schottky mixer has a peak SNR of 50 dB, although fewer lines are present than are visible from the HEB. As previously discussed, the difference between signals shown from two detectors mainly represents their differences in sensitivity, spectral response, and nonlinearity. In particular, saturation of the HEB generates several lines not present on the Schottky mixer, limiting its effective dynamic range to about 37 dB. Still, both detectors are clearly suitable for detecting strong multi-heterodyne signals.

\begin{figure}[H]
	\centering
	\includegraphics[width=0.81\linewidth]{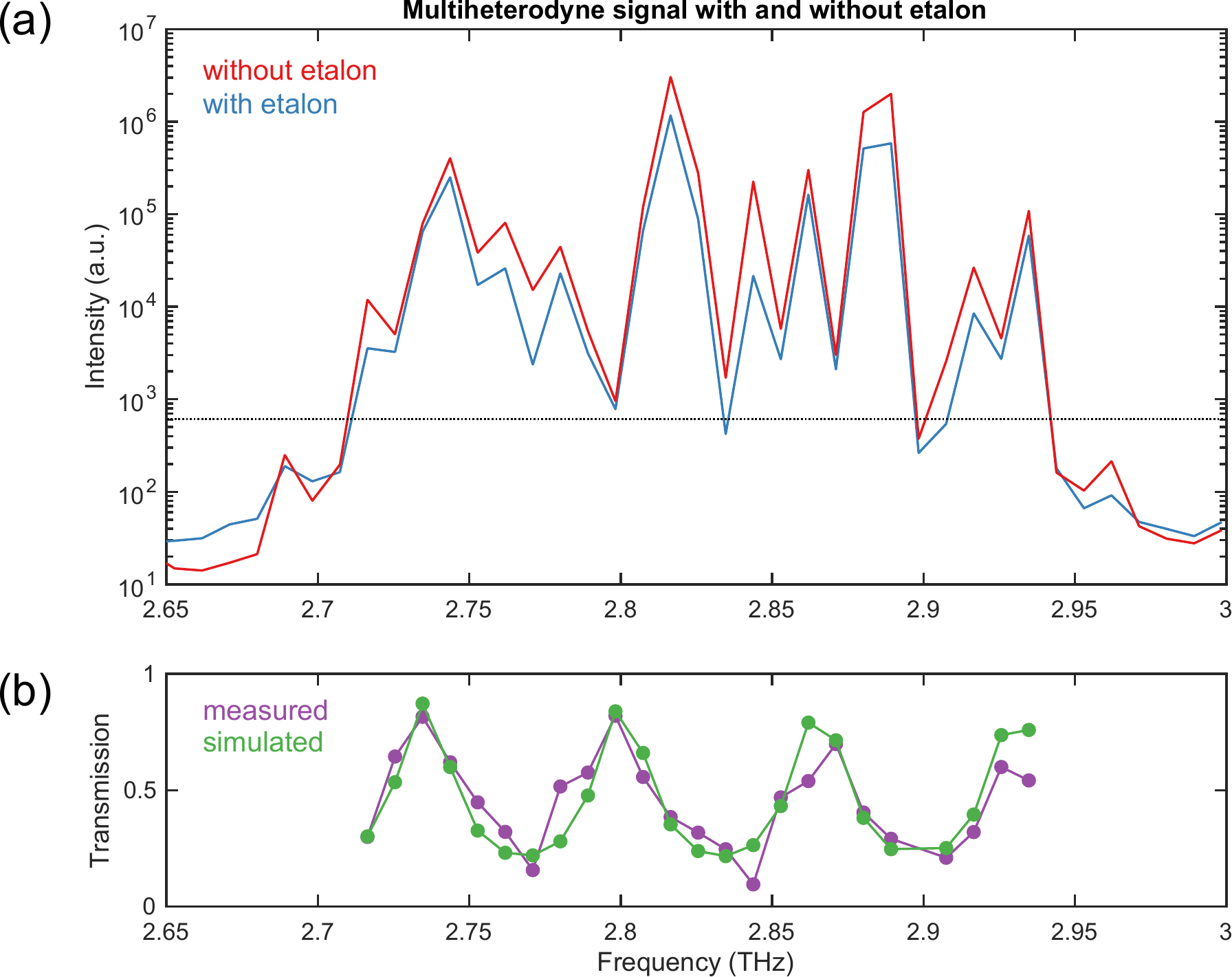}
	\caption{(a) Intensity of individual multi-heterodyne peaks with and without the GaAs etalon. The dashed horizontal line indicates the threshold for inclusion in the transmission data. (b) The measured etalon transmission (in purple) and the simulated etalon transmission (in green).}
	\label{fig:3}
\end{figure}

As a demonstration of broadband spectroscopy, we have performed transmission measurements of a low finesse etalon made from a tilted 625 $\mu$m-thick undoped GaAs wafer. The signal and LO lasers were shined onto the HEB, and the etalon was placed in the signal laser's path. For this measurement, no reference detector was used. Figure \ref{fig:3}(a) shows the multi-heterodyne data collected from the HEB with and without the etalon in blue and red, respectively. Figure \ref{fig:3}(b) shows the ratio of individual multi-heterodyne peaks along with the simulated transmission data at the frequencies that were sampled. To account for the dynamic range limitations of the HEB, we plotted only those transmission values corresponding to lines whose reference signal was greater than -37 dB from the peak intensity, indicated by the dashed horizontal line in Figure \ref{fig:3}(a). Periodic transmission due to the etalon is clearly visible within the lasing spectral range, and is in reasonable agreement with the theoretically-calculated etalon transmission. Because no reference detector was used, some residual errors were present, on account of relative intensity fluctuations that occurred between the two measurements with and without the sample.  

Lastly, we examine the feasibility of multi-heterodyne spectroscopy based on QCLs which are operated in pulsed biasing mode (not to be confused with the optical pulses of a mode-locked laser). It is well-known that operating QCLs in continuous-wave mode is significantly more challenging than operating the same devices in pulsed mode, because CW operation places much greater thermal constraints on the laser, in both the mid-infrared and the terahertz. Many gain media simply have thresholds that are too high for CW operation, and even when CW operation is possible the lasers' power dissipation becomes problematic. For dual comb THz spectroscopy, this is doubly problematic because the two lasers are placed inside the same cooler. In addition, for spectroscopy it is often desirable to have small repetition rates, as the dense mode spacing eases the constraints on the detector, and also makes it easier to achieve gapless coverage. Unfortunately, this necessarily entails longer lasers that consume more power. As an example, we constructed 7 mm combs from the gain medium of Ref. \cite{burghoff_terahertz_2014}, which consume approximately 1.3 A ($\sim 1000 \textrm{A/cm}^2$) and 15 V. Although these lasers have small free spectral ranges, around 4.8 GHz, two of them together consume approximately 40 W. This constitutes a major load on the cryocooler, and would result in the lasers warming to above their maximum CW operating temperature.

Traditionally, pulsed mode operation of QCLs is considered anathema to multi-heterodyne spectroscopy, as dual comb spectroscopy usually requires stable combs, while pulsed mode operation is inherently unstable. However, by using self-referenced SWIFTS \cite{Burghoff2015} to evaluate the coherence of similar devices, we have previously shown that even extremely unstable devices usually maintain their mutual coherence, at least in the most general sense that the modes remain evenly-spaced. This implies that by applying the same phase and timing correction techniques previously described, we should be able to recover multi-heterodyne lines even in the face of the large instability and chirp associated with pulsed operation. Although the absolute chirp of the lasers remains unknown, impeding the analysis of high-resolution features, it may alternatively be possible to exploit this effect to perform high resolution spectroscopy \cite{Nikodem2012}. 

\begin{figure}[H]
	\centering
	\includegraphics[width=\linewidth]{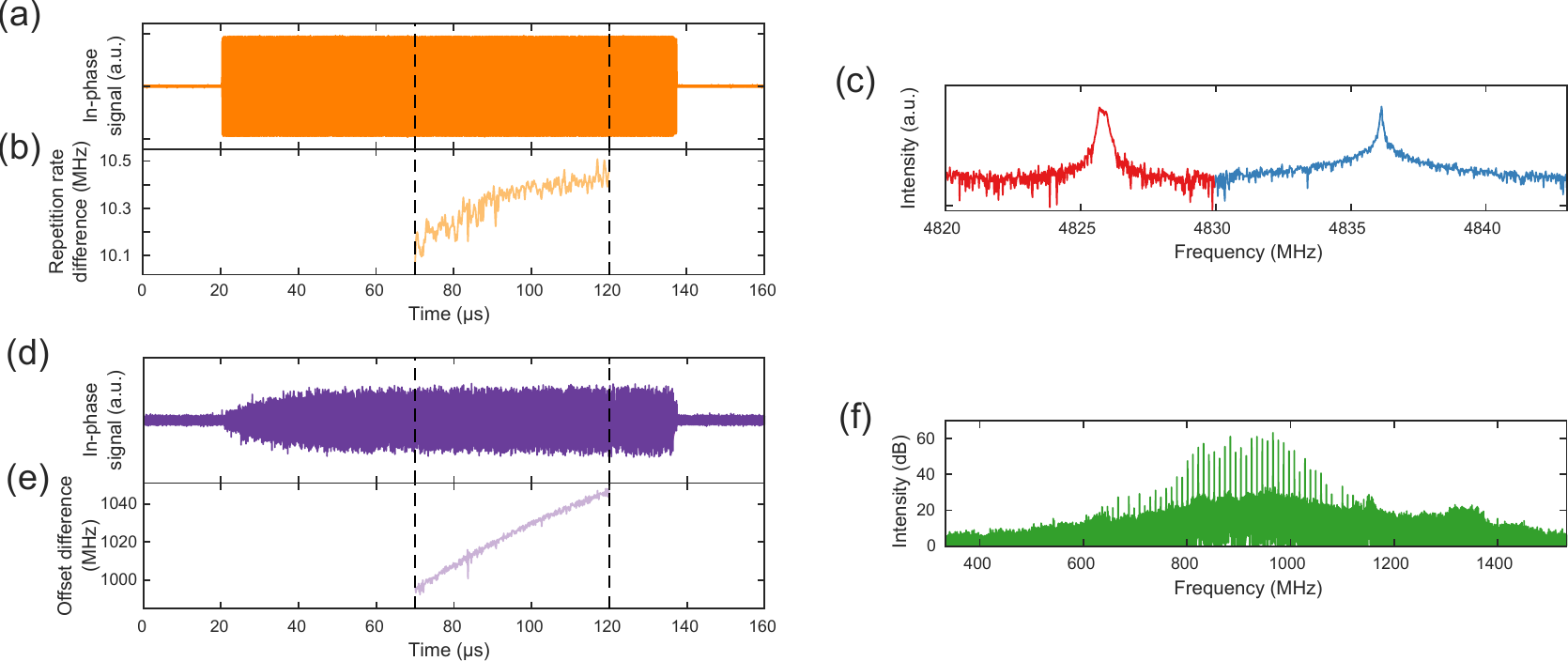}
	\caption{(a) Time domain signal of combs' repetition rate measured electrically from a bias tee. (b) Chirping of the repetition rates' difference in the 50 $\mu$s interval indicated by the dashed lines. (c) Two frequency combs' repetition rate beatnotes in the frequency domain, both of which are located around 4.8 GHz with a 10 MHz difference. (d) Time domain signal of the multi-heterodyne signal measured from the HEB detector. (e) Chirping of the multi-heterodyne signal's offset frequency in the 50 $\mu$s interval indicated by the dashed lines. (f) Multi-heterodyne signal in frequency domain. It is centered at 900 MHz with 45 observable modes, which corresponds to a 215 GHz coverage in the THz spectrum. }
	\label{fig:4}
\end{figure}

Figure \ref{fig:4} shows the results of pulsed mode dual comb spectroscopy using the aforementioned devices, whose gain media and dispersion compensation scheme are described in Ref. \cite{burghoff_terahertz_2014}. The lasers are biased to a comb regime using 120 $\mu$s pulses with a repetition frequency of 10 kHz, resulting in a duty cycle of 1.2\%. This low duty cycle significantly eases the cryogenic operation. A low pass filter was used to select only the part of the comb spectrum around 3.3 THz. Figures \ref{fig:4}(a) and \ref{fig:4}(d) respectively show in the time domain the combs' repetition rate signal (measured from a bias tee) and corresponding multi-heterodyne signal (measured from the HEB). As expected, both signals turn on during the electrical pulse, although the optical signal takes approximately 30 $\mu$s to stabilize. This reflects the fact that electrical beatnotes (which turn on within a few $\mu$s) are unreliable indicators of optical beatnotes. Figure \ref{fig:4}(c) shows the distinct repetition rate beatnotes in the frequency domain, clearly showing their frequency difference of 10 MHz. In pulsed mode, chirping of the repetition rate due to device heating is noticeable; this heating results in a substantial broadening of the beatnote indicated by red in Fig. \ref{fig:4}(c). (The lasers have slightly different beatnotes due to differences in the dispersion compensation and lens mounting, and the one indicated by blue possesses less chirp.)

When the difference in the repetition rates is plotted in the time domain, as shown in Fig. \ref{fig:4} (b), the chirp due to heating is evident. During the 50 $\mu$s period indicated by the two dashed lines, the difference of the combs' repetition rates gradually increases from 10.1 MHz to 10.5 MHz . In addition, Fig. \ref{fig:4}(e) shows the chirp of the offset frequency difference during the same 50 $\mu$s; it too is up-chirped. However, its magnitude is much larger, over 40 MHz, which is approximately 100 times the repetition rate chirping. Lastly, Fig. \ref{fig:4}(f) shows the phase- and timing-corrected multi-heterodyne signal in the frequency domain. The repetition rate difference of 10 MHz is clearly visible here, and over 45 modes are contributing to the multi-heterodyne signal, implying a coverage of 215 GHz in the THz spectrum. Within 50 $\mu$s of integration, the peak SNR of the multi-heterodyne signal is higher than 60 dB on the HEB. Of course, co-averaging multiple pulses further boosts the SNR.

In conclusion, we have demonstrated the first THz multi-heterodyne spectroscopy using quantum cascade laser frequency combs. The spectral coverage is 250 GHz at 2.8 THz, limited by the bandwidth of the lasing spectrum, and as a proof-of-principle we have used them to measure the transmission of a sample.  As broadband gain media designs continue to develop, we expect much broader coverage of the THz spectrum. Moreover, we have shown that even under pulsed mode operation, laser frequency combs are still feasible for multi-heterodyne spectroscopy. Pulsed mode multi-heterodyne spectroscopy shows great promise for reducing the cooling power constraints, allowing for the use of compact Stirling cryocoolers. Together with room temperature Schottky mixers as detectors, this will enable a very compact spectroscopy system at the THz region.


\section*{Acknowledgments}
The authors would like to acknowledge Jeffrey Hesler and Berhanu Bulcha from Virginia Diodes Inc. for the use of a Schottky mixer.

The work at MIT was supported by the DARPA SCOUT program through grant number W31P4Q-16-1-0001 from AMRDEC and NSF. The work in the Netherlands was supported by NWO, NATO SFP, and RadioNet. This work was performed, in part, at the Center for Integrated Nanotechnologies, a U.S. Department of Energy, Office of Basic Energy Sciences user facility. Sandia National Laboratories is a multi-program laboratory managed and operated by Sandia Corporation, a wholly owned subsidiary of Lockheed Martin Corporation, for the U.S. Department of Energy’s National Nuclear Security Administration under Contract No. DE-AC04-94AL85000.


\bibliographystyle{osajnlnt}
\bibliography{combref}

\begin{thebibliography}{10}
\newcommand{\enquote}[1]{``#1''}

\bibitem{Finneran2015}
I.~A. Finneran, J.~T. Good, D.~B. Holland, P.~B. Carroll, M.~A. Allodi, and
  G.~A. Blake, Physical review letters \textbf{114}, 163902 (2015).

\bibitem{Drouin2005}
B.~Drouin, F.~W. Maiwald, and J.~Pearson, Review of Scientific Instruments
  \textbf{76}, 093113 (2005).

\bibitem{Pearson2011}
J.~C. Pearson, B.~J. Drouin, A.~Maestrini, I.~Mehdi, J.~Ward, R.~H. Lin, S.~Yu,
  J.~J. Gill, B.~Thomas, C.~Lee, G.~Chattopadhyay, E.~Schlecht, F.~W. Maiwald,
  P.~F. Goldsmith, and P.~Siegel, Review of Scientific Instruments \textbf{82},
  1 (2011).

\bibitem{qin_tuning_2009}
Q.~Qin, B.~S. Williams, S.~Kumar, J.~L. Reno, and Q.~Hu, Nat Photon \textbf{3},
  732 (2009).

\bibitem{bartalini_frequency-comb-assisted_2014}
S.~Bartalini, L.~Consolino, P.~Cancio, P.~De~Natale, P.~Bartolini, A.~Taschin,
  M.~De~Pas, H.~Beere, D.~Ritchie, M.~S. Vitiello, and R.~Torre, Phys. Rev. X
  \textbf{4}, 021006 (2014).

\bibitem{ren_high-resolution_2011}
Y.~Ren, J.~N. Hovenier, R.~Higgins, J.~R. Gao, T.~M. Klapwijk, S.~C. Shi,
  B.~Klein, T.-Y. Kao, Q.~Hu, and J.~L. Reno, Applied Physics Letters
  \textbf{98}, 231109 (2011).

\bibitem{lu_widely_2014}
Q.~Y. Lu, S.~Slivken, N.~Bandyopadhyay, Y.~Bai, and M.~Razeghi, Applied Physics
  Letters \textbf{105}, 201102 (2014).

\bibitem{jiang_spectroscopic_2016}
Y.~Jiang, K.~Vijayraghavan, S.~Jung, A.~Jiang, J.~H. Kim, F.~Demmerle,
  G.~Boehm, M.~C. Amann, and M.~A. Belkin, Scientific Reports \textbf{6}, 21169
  (2016).

\bibitem{schiller_spectrometry_2002}
S.~Schiller, Optics Letters \textbf{27}, 766 (2002).

\bibitem{keilmann_time-domain_2004}
F.~Keilmann, C.~Gohle, and R.~Holzwarth, Optics Letters \textbf{29}, 1542
  (2004).

\bibitem{coddington_coherent_2008}
I.~Coddington, W.~C. Swann, and N.~R. Newbury, Phys. Rev. Lett. \textbf{100},
  013902 (2008).

\bibitem{newbury_sensitivity_2010}
N.~R. Newbury, I.~Coddington, and W.~Swann, Optics Express \textbf{18}, 7929
  (2010).

\bibitem{Tonouchi2007}
M.~Tonouchi, Nature Photonics \textbf{1}, 97 (2007).

\bibitem{Yasui2015}
T.~Yasui, R.~Ichikawa, Y.-D. Hsieh, K.~Hayashi, H.~Cahyadi, F.~Hindle,
  Y.~Sakaguchi, T.~Iwata, Y.~Mizutani, H.~Yamamoto, K.~Minoshima, and H.~Inaba,
  Scientific reports \textbf{5}, 10786 (2015).

\bibitem{Hugi2012}
A.~Hugi, G.~Villares, S.~Blaser, H.~C. Liu, and J.~Faist, Nature \textbf{492},
  229 (2012).

\bibitem{Khurgin2014}
J.~B. Khurgin, Y.~Dikmelik, A.~Hugi, and J.~Faist, Applied Physics Letters
  \textbf{104}, 081118 (2014).

\bibitem{burghoff_terahertz_2014}
D.~Burghoff, T.-Y. Kao, N.~Han, C.~W.~I. Chan, X.~Cai, Y.~Yang, D.~J. Hayton,
  J.-R. Gao, J.~L. Reno, and Q.~Hu, Nat Photon \textbf{8}, 462 (2014).

\bibitem{Wienold2014}
M.~Wienold, B.~R{\"{o}}ben, L.~Schrottke, and H.~T. Grahn, Optics Express
  \textbf{22}, 30410 (2014).

\bibitem{rosch_octave-spanning_2015}
M.~Rösch, G.~Scalari, M.~Beck, and J.~Faist, Nat Photon \textbf{9}, 42 (2015).

\bibitem{Li2015}
H.~Li, P.~Laffaille, D.~Gacemi, M.~Apfel, C.~Sirtori, J.~Leonardon,
  G.~Santarelli, M.~R{\"{o}}sch, G.~Scalari, M.~Beck, J.~Faist,
  W.~H{\"{a}}nsel, R.~Holzwarth, and S.~Barbieri, Optics express \textbf{23},
  33270 (2015).

\bibitem{wang_high-resolution_2014}
Y.~Wang, M.~G. Soskind, W.~Wang, and G.~Wysocki, Applied Physics Letters
  \textbf{104}, 031114 (2014).

\bibitem{villares_dual-comb_2014}
G.~Villares, A.~Hugi, S.~Blaser, and J.~Faist, Nat Commun \textbf{5}, 5192
  (2014).

\bibitem{Zhang2010}
W.~Zhang, P.~Khosropanah, J.~R. Gao, E.~L. Kollberg, K.~S. Yngvesson,
  T.~Bansal, R.~Barends, and T.~M. Klapwijk, Applied Physics Letters
  \textbf{96}, 111113 (2010).

\bibitem{itqw}
Y.~Yang, \enquote{Towards thz dual-comb spectrometer based on quantum cascade
  laser frequency combs,}  (2015).

\bibitem{Roy2012}
J.~Roy, J.-D. Desch{\^{e}}nes, S.~Potvin, and J.~Genest, Optics Express
  \textbf{20}, 21932 (2012).

\bibitem{ideguchi_adaptive_2014}
T.~Ideguchi, A.~Poisson, G.~Guelachvili, N.~Picqué, and T.~W. Hänsch, Nat
  Commun \textbf{5}, 3375 (2014).

\bibitem{Vitiello2012}
M.~S. Vitiello, L.~Consolino, S.~Bartalini, A.~Taschin, A.~Tredicucci,
  M.~Inguscio, and P.~{De Natale}, Nature Photonics \textbf{6}, 525 (2012).

\bibitem{Bartalini2010}
S.~Bartalini, S.~Borri, P.~Cancio, A.~Castrillo, I.~Galli, G.~Giusfredi,
  D.~Mazzotti, L.~Gianfrani, and P.~{De Natale}, Physical review letters
  \textbf{104}, 083904 (2010).

\bibitem{Burghoff2015}
D.~Burghoff, Y.~Yang, D.~J. Hayton, J.-R. Gao, J.~L. Reno, and Q.~Hu, Optics
  Express \textbf{23}, 1190 (2015).

\bibitem{Nikodem2012}
M.~Nikodem and G.~Wysocki, Sensors \textbf{12}, 16466 (2012).

\end{thebibliography}
\end{document}